\documentclass[aps,prl,preprint,russian,nofootinbib,showpacs]{revtex4}
\usepackage{graphicx}

\begin{document}

\title{Temperature dependence of the upper critical field in disordered
Hubbard model with attraction}

\author{E.Z. Kuchinskii$^1$, N.A. Kuleeva$^1$, M.V. Sadovskii$^1$$^,$$^2$}

\affiliation{$^1$Institute for Electrophysics, Russian Academy of Sciences, Ural Branch, 
Ekaterinburg 620016, Russia\\
$^2$M.N. Mikheev Institute for Metal Physics, Russian Academy of Sciences, Ural Branch, 
Ekaterinburg 620108, Russia}


\begin{abstract}

We study disorder effects upon the temperature behavior of the upper critical
magnetic field in attractive Hubbard model within the generalized DMFT+$\Sigma$ 
approach. We consider the wide range of attraction potentials $U$ -- from the
weak coupling limit, where superconductivity is described by BCS model, up to
the strong coupling limit, where superconducting transition is related to
Bose -- Einstein condensation (BEC) of compact Cooper pairs, formed at 
temperatures significantly higher than superconducting transition temperature,
as well as the wide range of disorder -- from weak to strong, when the system
is in the vicinity of Anderson transition. The growth of coupling strength
leads to the rapid growth of $H_{c2}(T)$, especially at low temperatures.
In BEC limit and in the region of BCS -- BEC crossover $H_{c2}(T)$ dependence
becomes practically linear. Disordering also leads to the general growth of
$H_{c2}(T)$. In BCS limit of weak coupling increasing disorder lead both to the
growth of the slope of the upper critical field in the vicinity of transition
point and to the increase of $H_{c2}(T)$ in low temperature region.
In the limit of strong disorder in the vicinity of the Anderson transition
localization corrections lead to the additional growth of $H_{c2}(T)$ at low
temperatures, so that the $H_{c2}(T)$ dependence becomes concave.
In BCS -- BEC crossover region and in BEC limit disorder only slightly influences
the slope of the upper critical field close to $T_{c}$. However, in the low
temperature region $H_{c2}(T)$ may significantly grow with disorder in the
vicinity of the Anderson transition, where localization corrections notably
increase $H_{c2}(T=0)$ also making $H_{c2}(T)$ dependence concave.

\end{abstract}


\maketitle


\section{Introduction}

The studies of disorder influence on superconductivity have rather long history.
In pioneer papers by Abrikosov and Gor'kov \cite{AG_impr,AG_imp,Gor_GL,AG_mimp} 
they analyzed the limit of weak disorder ($p_Fl\gg 1$, where $p_F$ is the Fermi
momentum and $l$ is the mean free path) and weak coupling superconductivity, 
which is well described by BCS theory. The well known ``Anderson theorem'' on
the critical temperature $T_c$ of superconductors with ``normal'' (nonmagnetic)
disorder \cite{And_th,Genn} is usually also attributed to this limit.

The generalization of the theory of ``dirty'' superconductors for the case of
strong enough disorder ($p_Fl\sim1$) (and up to the region of Anderson transition) 
was done in Refs. \cite{SCLoc_1,SCLoc_2,SCLoc_3,SCLoc_4}, where superconductivity 
was also analyzed in the weak coupling limit.

Most dramatically the  effects of disordering are reflected in the behavior of 
the upper critical magnetic field. In the theory of ``dirty'' superconductors
the growth of disorder leads to the increase both of the slope of the temperature
dependence of the upper critical field at $T_{c}$ \cite{Genn} and of $H_{c2}(T)$ 
in the whole temperature region \cite{WHH}. The effects of Anderson localization 
in the limit of strong enough disorder are also mostly reflected in the 
temperature dependence of the upper critical field. At the point of Anderson 
metal -- insulator transition transition itself, localization effects lead to
rather sharp increase of  $H_{c2}$ at low temperatures and temperature dependence
of $H_{c2}(T)$ is qualitatively different from the dependence derived by
Werthamer, Helfand and Hohenberg (WHH) \cite{WHH}, which is characteristic for the
theory of ``dirty'' superconductors and $H_{c2}(T)$ dependence becomes concave, i.e. 
demonstrates the positive curvature \cite{SCLoc_1,SCLoc_2,SCLoc_3}.

The problem of the generalization of BCS theory into the strong coupling region is
known for pretty long time. Significant progress in this direction was achieved in
paper by Nozieres and Schmitt -- Rink \cite{NS}, who proposed an effective method
to study the crossover from BCS -- like behavior in the weak coupling region
towards Bose -- Einstein condensation (BEC) in the strong coupling region.
At the same time, the problem of superconductivity of disordered systems in the
limit of strong coupling and in BCS -- BEC crossover region is still rather poorly
developed.

One of the simplest models to study BCS -- BEC crossover is Hubbard model with
attractive interaction. Most successful approach to the Hubbard model, both to
describe strongly correlated systems in the case of repulsive interaction, as well
as to study the BCS -- BEC crossover for the case of attraction, is the dynamical
mean field theory (DMFT) \cite{pruschke,georges96,Vollh10}.

In recent years we have developed the generalized DMFT+$\Sigma$ approach to the
Hubbard model \cite{JTL05,PRB05,FNT06,UFN12,HubDis,LVK16}, which is very convenient
for the studies of different external (with respect to those accounted by DMFT) 
interactions. In particular, this approach is well suited for the analysis of
two -- particle properties, such as optical (dynamic) conductivity \cite{HubDis,PRB07}.

In Ref. \cite{JETP14} we have used this approach to analyze single -- particle
properties of the normal phase and optical conductivity in attractive Hubbard model.
This was followed by our use of DMFT+$\Sigma$ in Ref. \cite{JTL14} to study
disorder influence on the temperature of superconducting transition, which was
calculated within Nozieres -- Schmitt-Rink approach. In particular, in this work
for the case of semi -- elliptic ``bare'' density of states (adequate for three -- 
dimensional case) we have numerically demonstrated the validity of the generalized
Anderson theorem, so that all effects of disordering on the critical temperature
(for all values of interaction parameter) are related only to general widening
of the ``bare'' band (density of states) by disorder.

An analytic proof of this universality of disorder influence on all single -- particle
properties in DMFT+$\Sigma$ approximation and on superconducting critical temperature
for the case of semi -- elliptic band was given in Ref. \cite{JETP15}.

Starting with classic work by Gor'kov \cite{Gor_GL}, it is well known that
Ginzburg -- Landau expansion is of fundamental importance for the theory of
``dirty'' superconductors, allowing the effective studies of the behavior of
various physical parameters close to superconducting critical temperature 
for different disorder levels \cite{Genn}. The generalization of this theory
(for weak coupling superconductors) to the region of strong disorder (up to
the Anderson metal -- insulator transition) was done in 
Refs. \cite{SCLoc_1,SCLoc_2,SCLoc_3}.

In Refs. \cite{JETP16,FNT16,JETP17} combining Nozieres -- Schmitt-Rink approximation
with DMFT+$\Sigma$ for attractive Hubbard models we provided microscopic derivation
of the coefficients of Ginzburg -- Landau expansion taking into account disordering,
which allowed the generalization of Ginzburg -- Landau theory to BCS -- BEC crossover
region and BEC limit of very strong coupling for different levels of disorder.
In particular, in Ref\cite{JETP17} using the generalization of self -- consistent
theory of localization this approach was extended to the case of strong disorder,
where Anderson localization effects become important. It was shown, that in the weak
coupling limit the slope of the $H_{c2}(T)$ dependence at $T=T_{c}$ increases with
disordering in the region of weak disorder in accordance with the theory of ``dirty''
superconductors, while in the limit of strong disorder localization effects lead to
the additional increase of the slope of the upper critical field. However, in the
region of BCS -- BEC crossover and in BEC limit the slope of $H_{c2}(T)$  close to
$T_{c}$ only slightly increases with the growth of disorder and the account of
localization effects is more or less irrelevant.

In the present paper, using the combination of Nozieres -- Schmitt-Rink and
DMFT+$\Sigma$ approximations for the attractive Hubbard model we shall analyze
disorder effects on the complete temperature dependence of $H_{c2}(T)$ for the
wide range of $U$ interaction values, including the region of BCS -- BEC crossover,
and the wide range of disorder levels up to the vicinity of the Anderson transition.

\section{Hubbard model within DMFT+$\Sigma$ approach in Nozieres -- Schmitt-Rink
approximation}

We consider the disordered nonmagnetic Anderson -- Hubbard model with attraction
described by the Hamiltonian:
\begin{equation}
H=-t\sum_{\langle ij\rangle \sigma }a_{i\sigma }^{\dagger }a_{j\sigma
}+\sum_{i\sigma }\epsilon _{i}n_{i\sigma }-U\sum_{i}n_{i\uparrow
}n_{i\downarrow },  
\label{And_Hubb}
\end{equation}
where $t>0$ is transfer integral between nearest neighbors, $U$ is the
Hubbard attraction on the lattice site, 
$n_{i\sigma }=a_{i\sigma }^{\dagger }a_{i\sigma }^{{\phantom{\dagger}}}$ --
number of electrons operator on the site, $a_{i\sigma }$ ($a_{i\sigma }^{\dagger}$) -- 
annihilation (creation) operator for an electron with spin $\sigma$, 
and local energies $\epsilon _{i}$ are assumed to be independent random variables
on different lattice sites. For the validity of the standard ``impurity'' diagram
technique \cite{Diagr,AGD} we assume the Gaussian distribution for energy levels
$\epsilon _{i}$:
\begin{equation}
\mathcal{P}(\epsilon _{i})=\frac{1}{\sqrt{2\pi}\Delta}\exp\left
(-\frac{\epsilon_{i}^2}{2\Delta^2}
\right)
\label{Gauss}
\end{equation}
Distribution width $\Delta$ serves as a measure of disorder and the Gaussian
random field of energy levels (independent on different lattice sites -- ``white
noise'' correlations) induces the ``impurity'' scattering, which is considered
within the standard approach, based on calculations of the averaged Green's
functions \cite{Diagr}.

The generalized DMFT+$\Sigma$ approach \cite{JTL05,PRB05,FNT06,UFN12} extends
the standard dynamical mean field theory (DMFT) \cite{pruschke,georges96,Vollh10}  
by addition of an ``external'' self -- energy (SE) $\Sigma_{\bf p}(\varepsilon)$
(in general case momentum dependent), which is related to any interaction outside
the limits of DMFT, and provides an effective method of calculations for both
single -- particle and two -- particle properties \cite{HubDis,PRB07}.
It completely conserves the standard self -- consistent equations of DMFT
\cite{pruschke,georges96,Vollh10}, while at each step of DMFT iteration procedure
the external SE $\Sigma_{\bf p}(\varepsilon)$ is recalculated again using some
approximate scheme, corresponding to the form of an external interaction and
the local Green's function of DMFT is also ``dressed'' by
$\Sigma_{\bf p}(\varepsilon)$ at each stage of the standard DMFT procedure.

In our problem of scattering by disorder \cite{HubDis,LVK16} for the ``external''
SE, entering DMFT+$\Sigma$ cycle, we use the simplest (self -- consistent Born)
approximation neglecting ``crossing'' diagrams for impurity scattering.
This ``external'' SE remains momentum independent (local).

To solve the effective single -- impurity Anderson model of DMFT in this paper, 
as in our previous works,  we use the very efficient method of numerical 
renormalization group (NRG) \cite{NRGrev}.

In the following we assume the ``bare'' band with semi -- elliptic density
of states (per unit cell with lattice parameter $a$ and for single spin
projection), which is reasonable approximation for three -- dimensional
case:
\begin{equation}
N_0(\varepsilon)=\frac{2}{\pi D^2}\sqrt{D^2-\varepsilon^2},
\label{DOSd3}
\end{equation}
where $D$ defines conduction band half -- width.

In Ref. \cite{JETP15} we have shown that in DMFT+$\Sigma$ approach for the
model with semi -- elliptic density of states all the influence of disorder on
single -- particle properties reduces simply to disorder induced band widening,
i.e. to the replacement $D\to D_{eff}$, where $D_{eff}$ is the effective band
half -- width of conduction band in the absence of correlations ($U=0$), 
widened by disorder:
\begin{equation}
D_{eff}=D\sqrt{1+4\frac{\Delta^2}{D^2}}.
\label{Deff}
\end{equation}
The ``bare'' (in the absence of $U$) density of states, ``dressed'' by disorder,
\begin{equation}
\tilde N_{0}(\xi)=\frac{2}{\pi D_{eff}^2}\sqrt{D_{eff}^2-\varepsilon^2},
\label{tildeDOS}
\end{equation}
remains semi -- elliptic also in the presence of disorder.

It should be noted that in other models of ``bare'' band disorder induces not
only widening of the band, but also changes the form of the density of states.
Thus, in general case there will be no complete universality of disorder influence 
on single -- particle properties, which is reduced to the simple replacement
$D\to D_{eff}$. However in the limit of strong disorder, which is of the main interest
to us, the ``bare'' band always becomes in practice semi -- elliptic and the
universality is restored \cite{JETP15}.

All calculations in this work, as in the previous, were done for rather typical
case of quarter -- filled band (number of electrons per lattice site $n$=0.5).

To analyze superconductivity for a wide range of pairing interaction $U$, 
following Refs. \cite{JETP14,JETP15}, we use Nozieres -- Schmitt-Rink approximation
\cite{NS}, which allows qualitatively correct (though approximate) description of
BCS -- BEC crossover region. In this approach, to determine the critical
temperature $T_c$ we use \cite{JETP15} the usual BCS -- like equation:
\begin{equation}
1=\frac{U}{2}\int_{-\infty}^{\infty}d\varepsilon \tilde N_0(\varepsilon)
\frac{th\frac{\varepsilon -\mu}{2T_c}}{\varepsilon -\mu},
\label{BCS}
\end{equation}
where the chemical potential $\mu$ for different values of $U$ and $\Delta$
is determined from DMFT+$\Sigma$ -- calculations, i.e. from the standard equation
for the number of electrons (band filling), which allows to find $T_c$  for the 
wide interval of model parameters, including the BCS -- BEC crossover region and
the limit of strong coupling, as well as for different levels of disorder.
This reflects the physical meaning of Nozieres -- Schmitt-Rink approximation:
in the weak coupling region transition temperature is controlled by the equation
for Cooper instability (\ref{BCS}), while in the limit of strong coupling it is
determined as BEC temperature controlled by chemical potential.

It was shown in Ref. \cite{JETP15}, that disorder influence on the critical
temperature $T_c$ and single -- particle characteristics (e.g. density of states)
in the model with semi -- elliptic density of states is universal and reduces only
to the change of the effective bandwidth.
In the weak coupling region the temperature of superconducting transition is well
described by BCS model, while in the strong coupling region the critical temperature
is mainly determined by the condition of Bose -- Einstein condensation of Cooper
pairs and decreases with the growth of $U$ as $t^2/U$, passing through a maximum at
$U/2D_{eff}\sim 1$. 

The review of this and similar results obtained for disordered Hubbard model in 
DMFT+$\Sigma$ approximation can be found in Ref. \cite{LVK16}.

\section{Basic relations for the upper critical field}

\begin{figure}
\includegraphics[clip=true,width=0.8\textwidth]{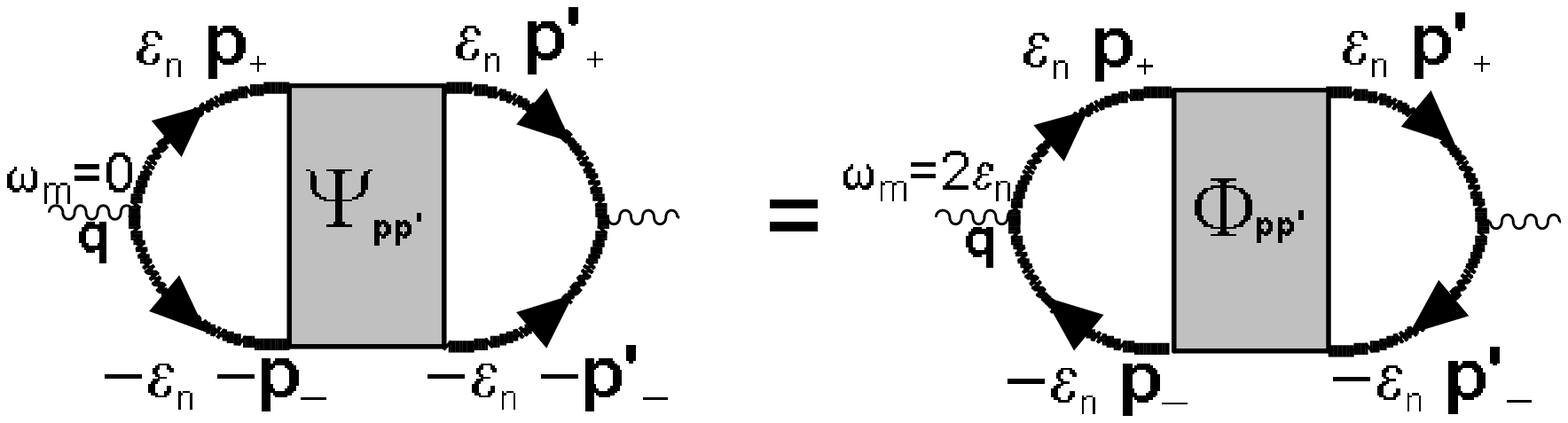}
\caption{Equivalence of loops in Cooper and diffusion channels in the case of
time reversal invariance.}
\label{diag_tinv}
\end{figure}

In Nozieres -- Schmitt-Rink approach the critical temperature of superconducting
transition is determined by combined solution of the weak coupling equation for 
Cooper instability in particle -- particle (Cooper) channel and the equation for
chemical potential for all values of Hubbard interaction within DMFT+$\Sigma$
procedure. The usual condition for Cooper instability is written as:
\begin{equation}
1=-U\chi({\bf q}),
\label{1l}
\end{equation}
where $\chi({\bf q})$ is Cooper susceptibility, determined by the loop diagram in 
Cooper channel, shown in Fig. \ref{diag_tinv}. In the presence of an external 
magnetic field total momentum in Cooper channel $\bf q$ acquires contribution from 
the vector potential $\bf A$
\begin{equation}
{\bf q}\to{\bf q}-\frac{2e}{c}{\bf A}.
\label{2l}
\end{equation}
As we assume isotropic electron spectrum, Cooper susceptibility $\chi({\bf q})$ 
depends on $\bf q$ only via $q^2$.
The minimal eigenvalue of ${({\bf q}-\frac{2e}{c}{\bf A})}^2$, determining
(orbital)\footnote{In this paper we do not consider paramagnetic effect due to
electronic spin.} upper critical magnetic field
$H=H_{c2}$ is given by \cite{LP}
\begin{equation}
{q_{0}}^2=2\pi\frac{H}{\Phi_{0}},
\label{3l}
\end{equation}
where $\Phi_{0}=\frac{ch}{2e}=\frac{\pi\hbar}{e}$ is magnetic flux quantum. 
Then the equation for $T_{c}(H)$ or $H_{c2}(T)$ remains the same:
\begin{equation}
1=-U\chi(q^2={q_0}^2).
\label{4l}
\end{equation}
In the following we shall neglect relatively weak magnetic field influence on
diffusion processes (broken time reversal invariance), which is reflected in non
equality of loop diagrams in Cooper and diffusion channels.
This influence was analyzed in Refs. \cite{SCLoc_3,SCLoc_4,Hc2loc_d3, Hc2loc_d3a},
where it was shown that the account of this broken symmetry only slightly decreases
the value of $H_{c2}(T)$ at low temperatures, even close to the Anderson transition.
In the case of time reversal invariance and due the static nature of impurity
scattering ``dressing'' two -- particle Green's function
$\Psi_{\bf p,\bf {p'}}( \varepsilon_n,{\bf q})$ we can change directions of all
lower electronic lines in the loop with simultaneous sign change of all momenta
on these lines (cf. Fig.\ref{diag_tinv}). Then we obtain:
\begin{equation}
\Psi_{\bf p,\bf {p'}}( \varepsilon_n,{\bf q})=
\Phi_{\bf p,\bf {p'}}(\omega_m =2\varepsilon_n,{\bf q}),
\label{l1}
\end{equation}
where $\varepsilon_n$ are Fermionic Matsubara frequencies,
${\bf p_{\pm}}={\bf p} \pm \frac{\bf q}{2}$, and
$\Phi_{\bf p,\bf {p'}}(\omega_m =2\varepsilon_n,{\bf q})$ is the two -- particle
Green's function in diffusion channel, dressed by impurities.
The we obtain the Cooper susceptibility as:
\begin{equation}
\chi({\bf q})=-T\sum_{n, \bf p, \bf {p'}}\Psi_{\bf p,\bf {p'}}
(\varepsilon_n,{\bf q})=-T\sum_{n, \bf p, \bf {p'}}\Phi_{\bf p,\bf {p'}}
(\omega_m =2\varepsilon_n,{\bf q}).
\label{l2}
\end{equation}

Performing the standard summation over Fermionic Matsubara frequencies
\cite{Diagr,AGD} we obtain for Cooper susceptibility entering Eq. (\ref{4l}):
\begin{equation}
\chi({q_0}^2)=
-\frac{1}{2\pi}\int_{-\infty}^{\infty}d\varepsilon
Im\Phi^{RA}(\omega =2\varepsilon,{q_0}^2)th\frac{\varepsilon}{2T},
\label{l3}
\end{equation}
where $\Phi^{RA}(\omega,{\bf q})=\sum_{\bf p, \bf {p'}}
\Phi_{\bf p, \bf {p'}}^{RA}(\omega,{\bf q})$.
To find the loop $\Phi^{RA}(\omega,{\bf q})$ in the case of strong disorder
(including the region of Anderson localization) we use the approximate self --
consistent theory of localization \cite{VW,WV,MS,KS,MS86,VW92,Diagr}. 
Then this loop contains the diffusion pole contribution, which is written as 
\cite{HubDis}:
\begin{equation}
\Phi^{RA}(\omega=2\varepsilon,{q_{0}}^2)=
-\frac{\sum_{\bf {p}}\Delta G_{\bf p}(\varepsilon)}
{\omega +iD(\omega){q_{0}}^2},
\label{l4}
\end{equation}
where $\Delta G_{\bf p}(\varepsilon)=G^{R}(\varepsilon,{\bf p})-
G^{A}(-\varepsilon,{\bf p})$, $G^{R}$ and $G^{A}$ are retarded and advanced Green's
functions, while $D(\omega)$ is frequency dependent generalized diffusion
coefficient. As a result, Eq. (\ref{4l}) for $H_{c2}(T)$ takes the form:
\begin{equation}
1=-\frac{U}{2\pi}\int_{-\infty}^{\infty}d\varepsilon
Im
\left(
\frac{\sum_{\bf {p}}\Delta G_{\bf p}(\varepsilon)}
{2\varepsilon+iD(2\varepsilon)2\pi\frac{H_{c2}}{\Phi_{0}}}
\right)
th\frac{\varepsilon}{2T}.
\label{l5}
\end{equation}

The generalized diffusion coefficient in self -- consistent theory of localization
\cite{VW,WV,MS,KS,MS86,VW92,Diagr} for the model under consideration is determined
by the following self -- consistence equation \cite{HubDis}:
\begin{equation}
D(\omega)=i\frac{<v>^2}{d}
\left(\omega-\Delta\Sigma_{imp}^{RA}(\omega)+\Delta^4\sum_{\bf p}
\Delta G_{\bf p}^{2}(\varepsilon)\sum_{\bf {q}}\frac{1}{\omega+iD(\omega)q^2}
\right)^{-1},
\label{l9}
\end{equation}
where $\omega=2\varepsilon$, 
$\Delta\Sigma_{imp}^{RA}(\omega)=\Sigma_{imp}^{R}(\varepsilon)-
\Sigma_{imp}^{A}(-\varepsilon)$, $d$ is space dimensionality, while the average
velocity $<v>$ is defined here as:
\begin{equation}
<v>=\frac{\sum_{\bf {p}}|{\bf {v_p}}|\Delta G_{\bf p}(\varepsilon)}
{\sum_{\bf {p}}\Delta G_{\bf p}(\varepsilon)}; 
{\bf {v_p}}=\frac{\partial\varepsilon (\bf p)}{\partial\bf p}.
\label{l7}
\end{equation}
Taking into account the limits of diffusion approximation
summation over $q$ in Eq. (\ref{l9}) should be limited by
\cite{MS86,Diagr}
\begin{equation}
q<k_0=Min \{l^{-1},p_F\},
\label{cutoff}
\end{equation}
where $l$ is the mean -- free path due to elastic scattering by disorder and
$p_F$ is  Fermi momentum.

In the limit of weak disorder, when localization corrections are small,
Cooper susceptibility $\chi({\bf q})$ is determined by ladder
approximation. In this approximation Cooper susceptibility
was studied by us in Ref. \cite{FNT16}.
Let us now rewrite self -- consistency Eq. (\ref{l9}) so that in the
limit of weak disorder it explicitly reproduces the results of ladder
approximation. In this approximation we neglect all contributions to
irreducible vertex from ``maximally crossed'' diagrams and the last term
in the r.h.s. of Eq. (\ref{l9}) just vanishes.
Now we introduce the frequency dependent generalized diffusion coefficient
in ladder approximation as:
\begin{equation}
D_{0}(\omega)=\frac{<v>^2}{d}
\frac{i}{\omega-\Delta\Sigma_{imp}^{RA}(\omega)}.
\label{l11}
\end{equation}
The value of $\frac{<v>^2}{d}$, entering the self -- consistency Eq. (\ref{l9}),
can now be expressed via this diffusion coefficient $D_{0}$ in ladder
approximation. Then the self -- consistency Eq. (\ref{l9}) takes the form:
\begin{equation}
D(\omega=2\varepsilon)=D_{0}(\omega=2\varepsilon)
\left(
1+\frac{\Delta ^4}
{2\varepsilon-\Delta\Sigma_{imp}^{RA}(\omega=2\varepsilon)}
\sum_{\bf p}\Delta G_{\bf p}^{2}(\varepsilon)
\sum_{\bf q}
\frac{1}{2\varepsilon+iD(\omega=2\varepsilon)q^2}
\right)^{-1}.
\label{l12}
\end{equation}
In the framework of the approach of Ref. \cite{FNT16} the diffusion coefficient
$D_{0}(\omega=2\varepsilon)$ in ladder approximation can be obtained in analytic
form. In fact, in the ladder approximation, the two -- particle Green's
function (\ref{l4}) can be written as:
\begin{equation}
\Phi_{0}^{RA}(\omega=2\varepsilon,{\bf q})=
-\frac{\sum_{\bf {p}}\Delta G_{\bf p}(\varepsilon)}
{\omega +iD_{0}(\omega=2\varepsilon)q^2}.
\label{l13}
\end{equation}
Let us introduce
\begin{equation}
\varphi(\varepsilon,{\bf q}=0)\equiv
\lim_{q \to 0}\frac{\Phi_{0}^{RA}(\omega=2\varepsilon,{\bf q})
-\Phi_{0}^{RA}(\omega=2\varepsilon,{\bf q}=0)}{q^2}=
\frac{i\sum_{\bf {p}}\Delta G_{\bf p}(\varepsilon)}{{\omega}^2}D_{0}
(\omega=2\varepsilon).
\label{l14}
\end{equation}
Then the diffusion coefficient $D_{0}$ can be written as:
\begin{equation}
D_{0}=\frac{\varphi(\varepsilon,{\bf q}=0)(2\varepsilon)^2}
{i\sum_{\bf {p}}\Delta G_{\bf p}(\varepsilon)}.
\label{l15}
\end{equation}
In Ref. \cite{FNT16}, using the exact Ward identity, written in ladder
approximation, it was shown that $\varphi(\varepsilon,{\bf q}=0)$
can be expressed as
\begin{equation}
\varphi(\varepsilon,{\bf q}=0)(2\varepsilon)^2=\sum_{\bf {p}}v_{x}^2
G^{R}(\varepsilon,{\bf p})G^{A}(-\varepsilon,{\bf p})+
\frac{1}{2}\sum_{\bf {p}}
\frac{\partial^2\varepsilon (\bf p)}{\partial p_{x}^2}
(G^{R}(\varepsilon,{\bf p})+G^{A}(-\varepsilon,{\bf p})),
\label{l16}
\end{equation}
where $v_x=\frac{\partial\varepsilon (\bf p)}{\partial p_{x}}$.

The procedure for the numerical now looks as follows.
First, using Eqs. (\ref{l16}), (\ref{l15}) we find the diffusion coefficient
$D_{0}$ in the ladder approximation. Then using self -- consistency 
Eq. (\ref{l12}) we find the generalized diffusion coefficient and solve 
Eq. (\ref{l5}) to determine  $H_{c2}(T)$.

\section{Main results}

\begin{figure}
\includegraphics[clip=true,width=0.5\textwidth]{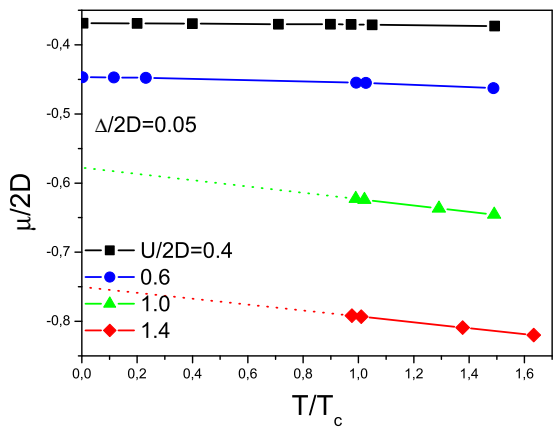}
\caption{Temperature dependence of the chemical potential for $\Delta/2D=0.05$ 
and different values of interaction.}
\label{fig2}
\end{figure}

The chemical potential enters Eq. (\ref{l5}) defining $H_{c2}(T)$ as a parameter,
which is to be determined from from the condition of band (quarter)filling using the
DMFT+$\Sigma$ procedure. Chemical potential depends not only on the coupling strength,
but also on the temperature, and this dependence is quite important in determining
the value of $H_{c2}(T)$ in the limit of strong enough coupling.
NRG algorithm we use as an impurity solver of DMFT neglects electronic levels
quantization in magnetic field, i.e. magnetic field influence on electron orbital
motion and correspondingly on the chemical potential. In Ref. \cite{JETP14} we have
shown, that in attractive Hubbard model our DMFT procedure becomes unstable for
$T<T_{c}$, which is reflected in finite difference of even and odd iterations of DMFT.
This instability is apparently related to instability of the normal state for
$T<T_{c}$. In particular, it is most sharp in BEC strong coupling limit
(for $U/2D \geq 1$), which makes impossible to determine the chemical potential
at $T<T_{c}$. In the weak coupling limit the difference between the results of
even and odd DMFT iterations is very small, which allows to find the values of
$\mu(T)$ with high accuracy even for $T<T_{c}$.
In Fig.\ref{fig2} we show the temperature dependence of the chemical potential
for different values of coupling strength. In the weak coupling limit
($U/2D=0.4,0.6$) in Fig.\ref{fig2} we show data obtained from DMFT+$\Sigma$
calculations, including the region of $T<T_{c}$.
In the limit of strong coupling we can determine the chemical potential directly
form DMFT+$\Sigma$ procedure only at $T>T_{c}$ and appropriate data points are
also shown in Fig. \ref{fig2}. From this figure we can see, that in the presence
of interactions the chemical potential acquires the linear temperature dependence,
which is quite important for us. In the weak coupling limit the chemical potential
does not have any singularities for $T<T_{c}$ and we can assume, that in the strong
coupling region $\mu(T)$ follows the same type of temperature dependence, which can
be found from linear extrapolation (dashed lines for $U/2D=1.0,1.4$ in Fig.\ref{fig2}) 
from the region of $T>T_{c}$. This procedure was used in our calculations for
for the strong coupling region.

\begin{figure}
\includegraphics[clip=true,width=0.9\textwidth]{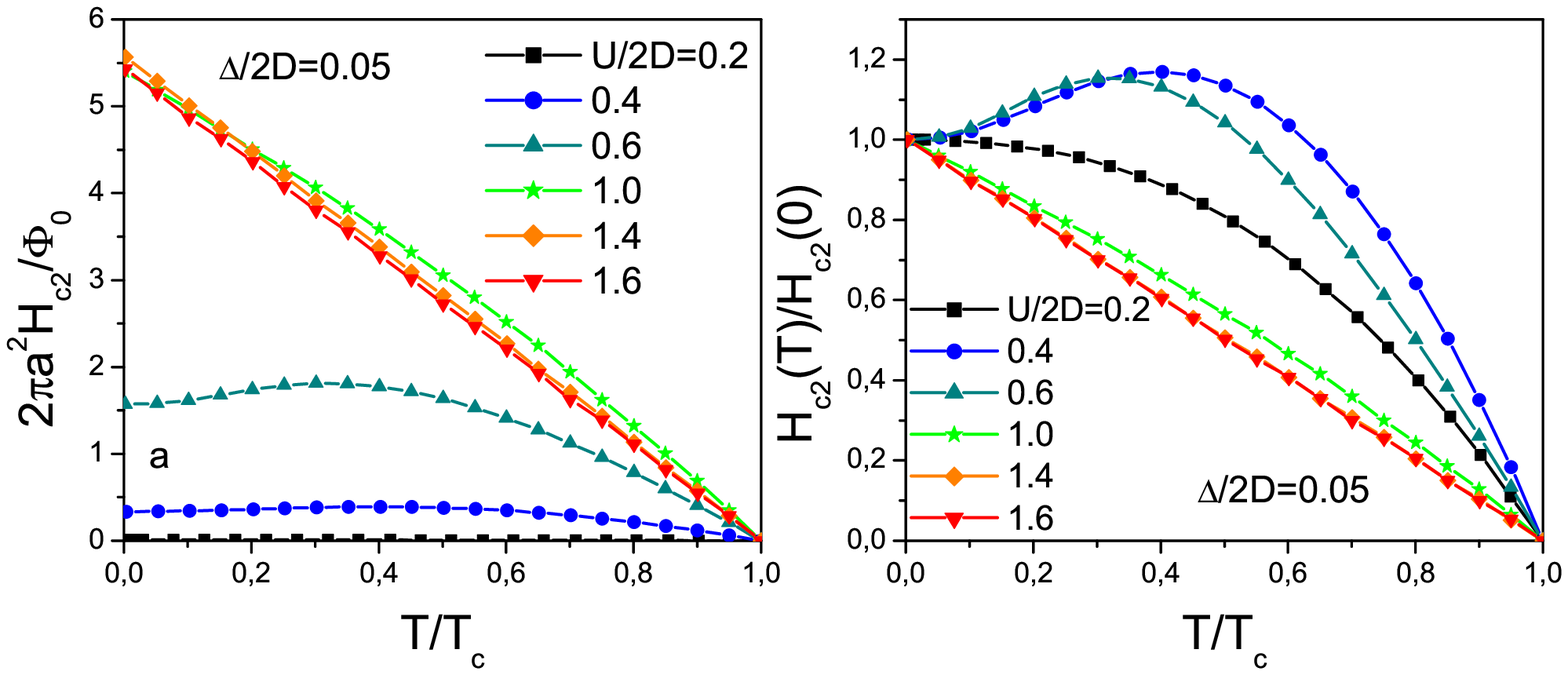}
\includegraphics[clip=true,width=0.9\textwidth]{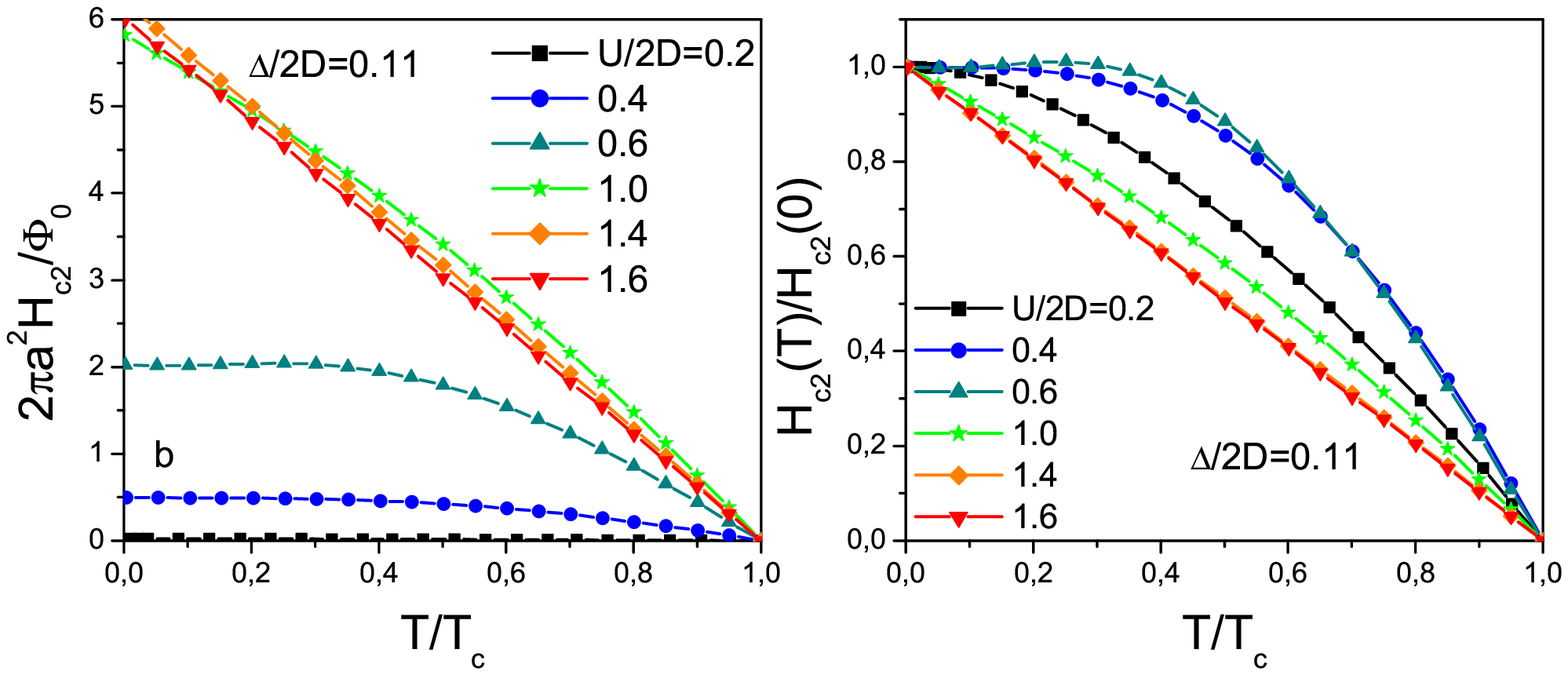}
\includegraphics[clip=true,width=0.9\textwidth]{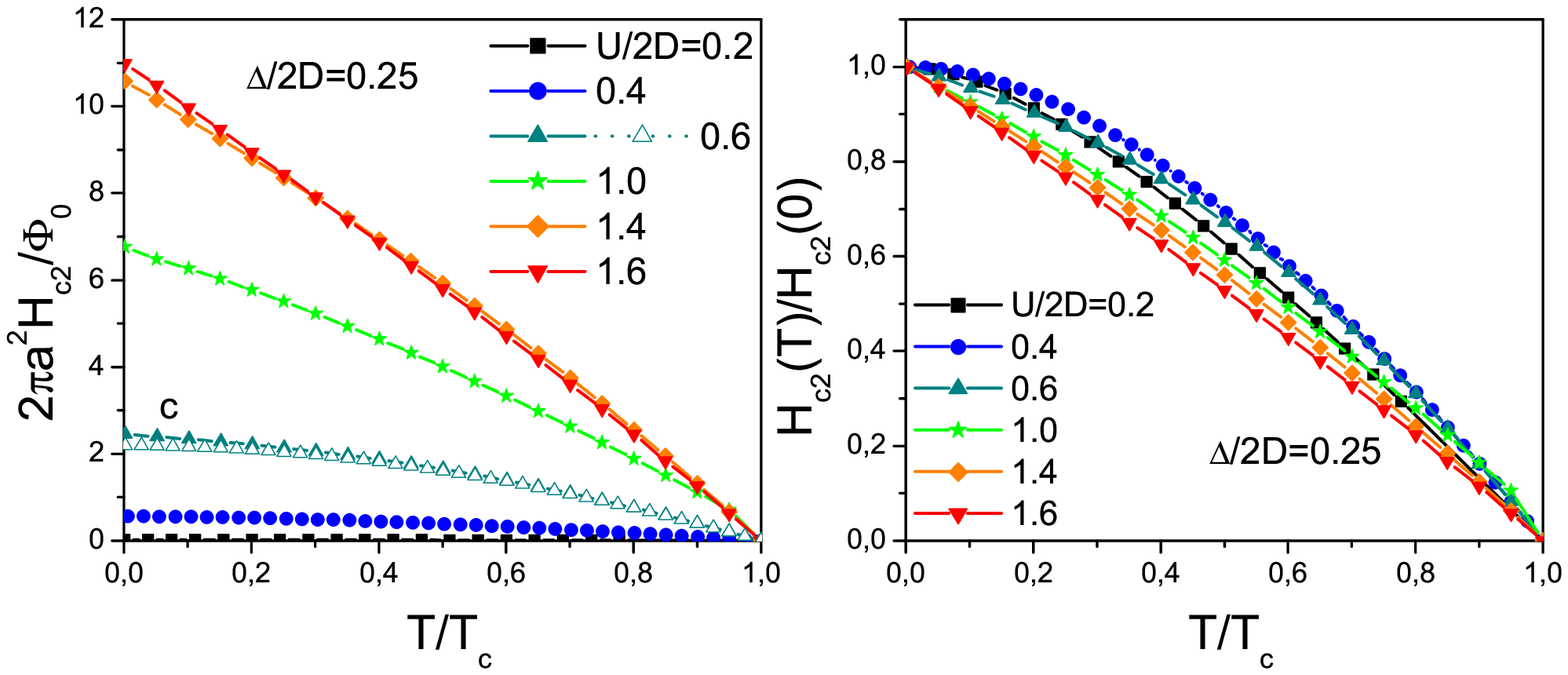}
\caption{
Temperature dependence of the upper critical field at different values of the
coupling strength for three different disorder levels:
(a) --- $\Delta/2D=0.05$;
(b) --- $\Delta/2D=0.11$;
(c) --- $\Delta/2D=0.25$.
On the left panels the upper critical field is normalized by
$\frac{\Phi_{0}}{2\pi a^2}$, while on right panels the upper critical field is 
normalized by its value at $T=0$.}
\label{fig3}
\end{figure}

In the limit of weak disorder ($\Delta/2D=0.05$ in Fig.\ref{fig3}(a)) and weak
coupling ($U/2D=0.2$) we observe the temperature dependence of the upper
critical field similar to the standard WHH dependence \cite{WHH} with negative
curvature. The growth of the coupling strength in general leads to significant 
increase of the upper critical field up to extremely high values over 
$\frac{\Phi_{0}}{2\pi a^2}$ ($a$ -- lattice spacing) in the low temperature
region. At intermediate couplings ($U/2D=0.4,0.6$) the temperature dependence
of $H_{c2}(T)$ acquires weak maximum at $T/T_{c}\sim (0.2-0.4)$.
Further increase of the coupling strength leads to the growth of the
upper critical field  and for $U/2D=1$ the temperature dependence $H_{c2}(T)$ 
becomes almost linear and for higher couplings the temperature dependence 
the value of the upper critical field remains practically the same for all
temperatures. With the growth of disorder ($\Delta/2D=0.11$ in Fig.\ref{fig3}(b)) 
situation remains qualitatively similar.
The increase of the coupling strength leads at first to the growth of $H_{c2}$ 
for all temperatures. The small maximum of $H_{c2}(T)$, observed at intermediate
couplings ($U/2D=0.4,0.6$) and weak disorder ($\Delta/2D=0.05$) vanishes.
In the strong coupling region ($U/2D \geq 1$) $H_{c2}(T)$ is in fact linear and 
only weakly changes with coupling strength. At strong enough disorder
($\Delta/2D=0.25$) with the growth of coupling strength the upper critical field
also grows in the whole temperature region. This growth continues up to BEC
region of very strong coupling ($U/2D=1.4$), after that $H_{c2}(T)$ dependence
becomes linear and only weakly dependent on the coupling strength.
For comparison on the left panel of Fig.\ref{fig3}(c) for $U/2D=0.6$ we show
both data obtained using self -- consistent theory of localization  
(filled triangles and continuous curve) and those calculated from ladder
approximation for impurity scattering (unfilled triangles and dashed curve). 
Weak difference between these dependencies demonstrates that corrections from
Anderson localization at this disorder level ($\Delta/2D=0.25$) are rather weak.

\begin{figure}
\includegraphics[clip=true,width=\textwidth]{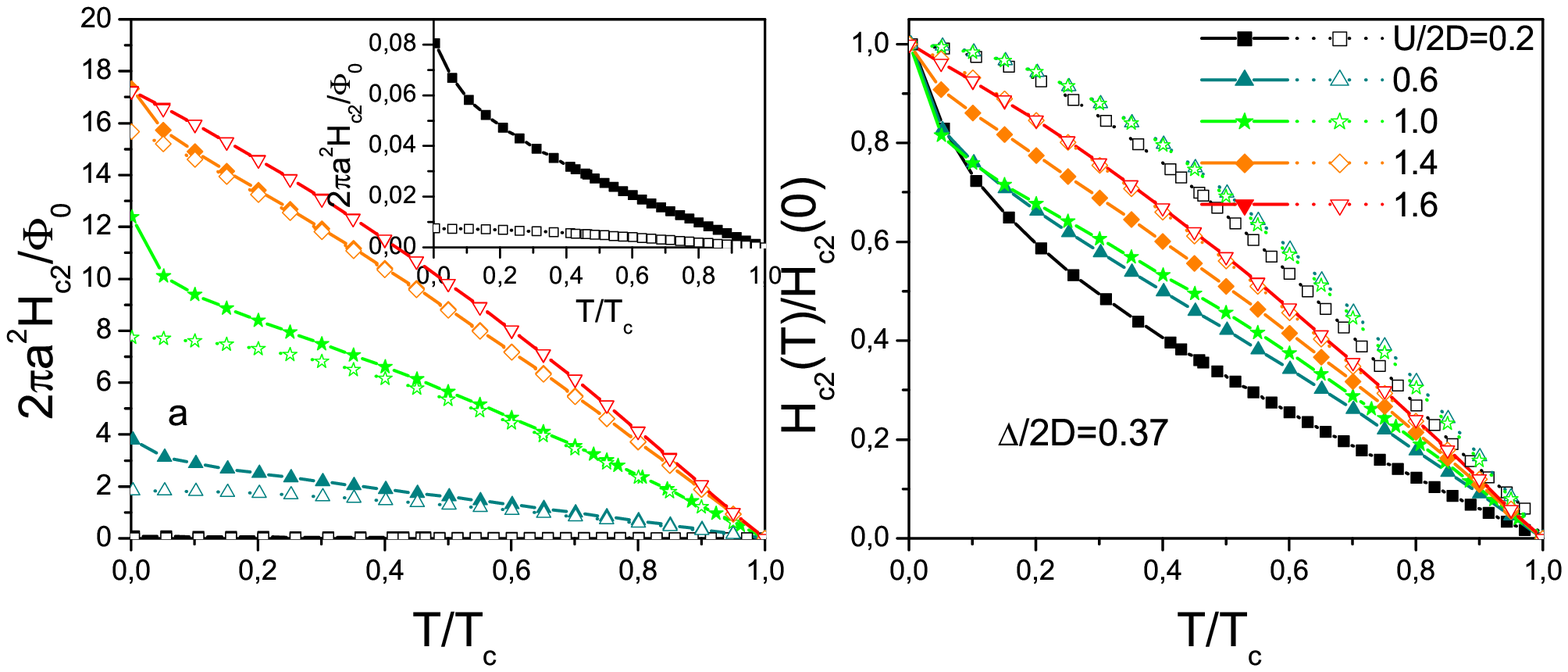}
\includegraphics[clip=true,width=\textwidth]{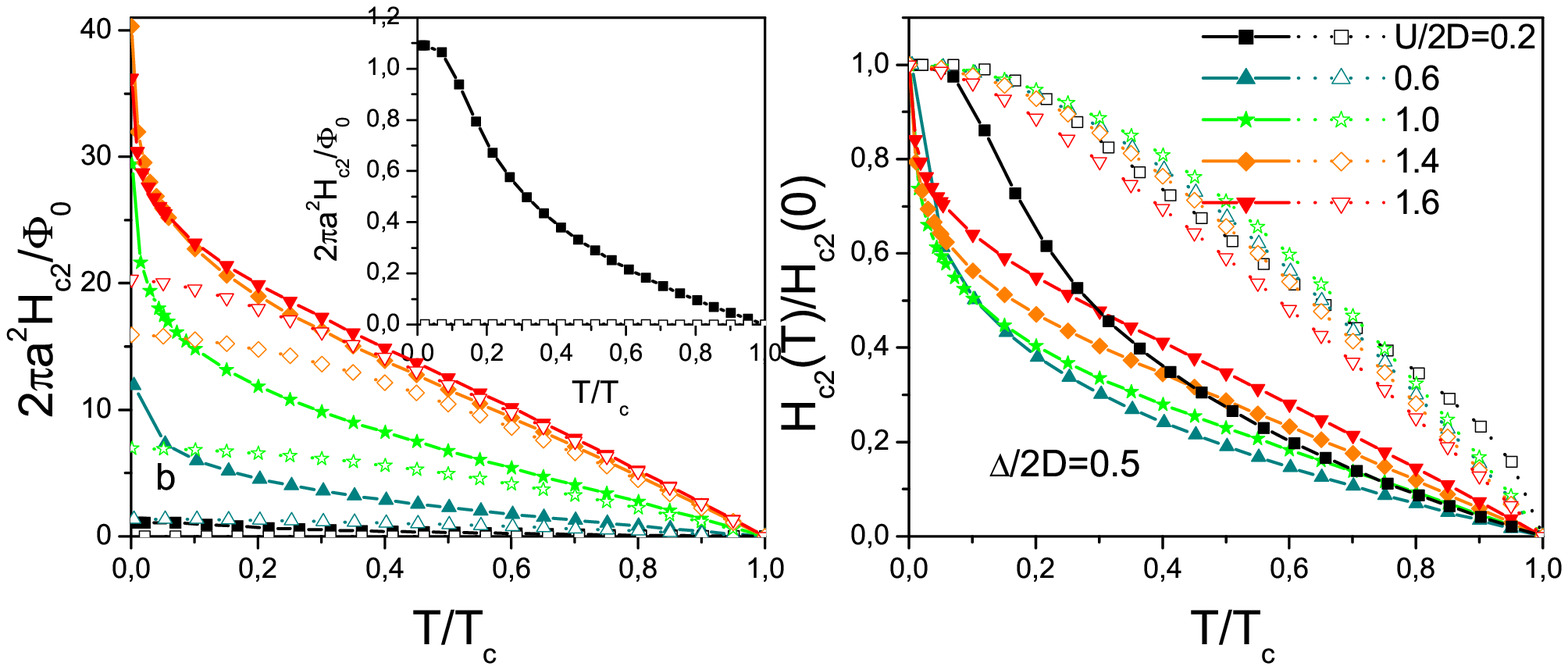}
\caption{
Temperature dependence of the upper critical field at Anderson metal -- insulator
transition ((a) --- $\Delta/2D=0.37$) and in Anderson insulator
((b) --- $\Delta/2D=0.5$) for different values of coupling strength.
On the left panels $H_{c2}$ is normalized by $\frac{\Phi_{0}}{2\pi a^2}$, on the 
right --- the upper critical field is normalized by its value at $T=0$}
\label{fig4}
\end{figure}

 In model under consideration in DMFT+$\Sigma$ approximation the Anderson
metal -- insulator transition occurs at $\Delta/2D=0.37$ and this value of 
critical disorder is independent of the coupling strength (cf.\cite{HubDis}). 
Temperature behavior of the upper critical field precisely at the point of
Anderson transition and in Anderson insulator phase for different values of
coupling strength is shown in Fig. \ref{fig4}. 
In this figure filled symbols and continuous curves show results of calculations
using the self -- consistent theory of localization, while unfilled symbols and
dashed curves correspond to the results of calculations using the ladder
approximation for impurity scattering. At the point of Anderson transition
($\Delta/2D=0.37$ in Fig. \ref{fig4}(a)) and in the limit of weak coupling 
localization effects strongly change the temperature dependence of $H_{c2}(T)$.
In particular these effects enhance $H_{c2}(T)$ in the whole temperature region. 
However, the greatest increase is observed at low temperatures, so that $H_{c2}(T)$
dependence acquires positive curvature, as was first shown in Refs.
\cite{SCLoc_1,SCLoc_2}. The increase of the coupling strength leads to the growth
of the upper critical field in the whole temperature interval. The curves of
$H_{c2}(T)$ in the intermediate coupling region ($U/2D=0.6,1$) still have 
positive curvature. Further increase of the coupling up to $U/2D=1.4$ also
enhance $H_{c2}$ at all temperatures. However, the account of localization
corrections at such a strong coupling is relevant only at low temperatures
($T/T_{c}<0.1$). In this region the $H_{c2}(T)$ dependence has positive
curvature, while at other temperatures $H_{c2}(T)$ is in fact linear. 
With further increase of coupling strength ($U/2D=1.6$) $H_{c2}(T)$ becomes
practically linear and localization correction become irrelevant at all
temperatures. Thus, in BEC limit of very strong coupling the influence of
of Anderson localization on the behavior of the upper critical field is
rather weak. In Anderson insulator phase (Fig.\ref{fig4}(b)) and in BCS weak
coupling limit ($U/2D=0.2$) the account of localization effects leads to
significant growth of $H_{c2}(T)$ (cf. insert in Fig.\ref{fig4}(b)). 
The increase of coupling strength leads to the growth of the upper critical
field in the whole temperature region. At intermediate couplings
($U/2D=0.6,1.0$) the account of localization effects notably increases
$H_{c2}$ for all temperatures.
However, the most significant increase is observed in the region of low
temperatures, leading to the positive curvature of $H_{c2}(T)$ dependence
and very sharp growth of $H_{c2}(T=0)$. In BEC limit of very strong coupling
($U/2D=1.4,1.6$) the upper critical field almost does not grow with coupling
strength. Contribution from localization effects for $T\sim T_{c}$ is
irrelevant and $H_{c2}(T)$ dependence is practically linear. However, at low
temperatures ($T\ll T_{c}$) contribution from Anderson localization still
significantly enhances the upper critical field and $H_{c2}(T)$ curve has
positive curvature. Thus, both in Anderson insulator phase and in BEC limit
of very strong coupling the influence of Anderson localization on the behavior
of the upper critical field is noticeably suppressed, though at low temperatures
it still remains quite significant changing the value of $H_{c2}(T=0)$.

\begin{figure}
\includegraphics[clip=true,width=0.5\textwidth]{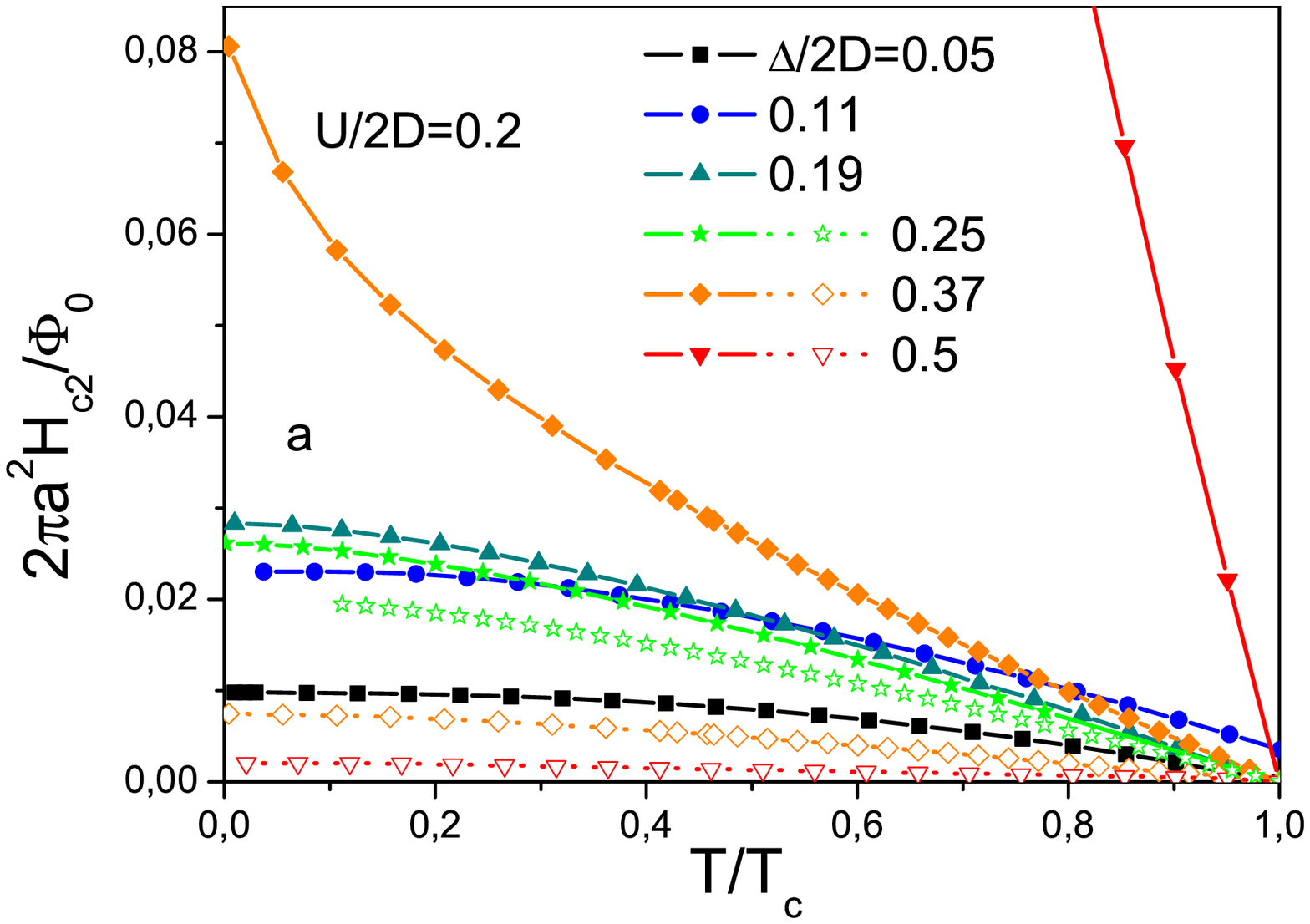}
\includegraphics[clip=true,width=0.5\textwidth]{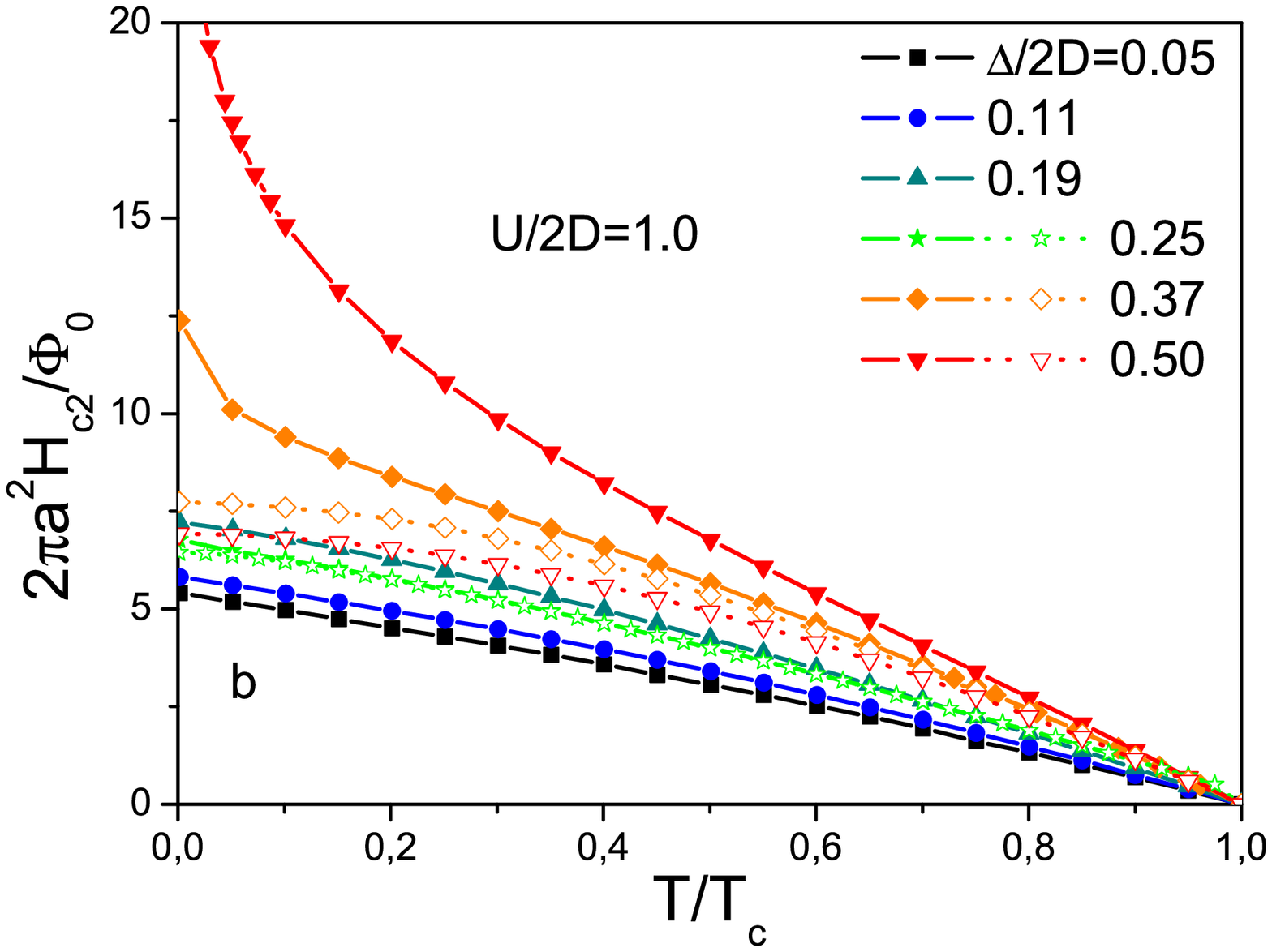}
\includegraphics[clip=true,width=0.5\textwidth]{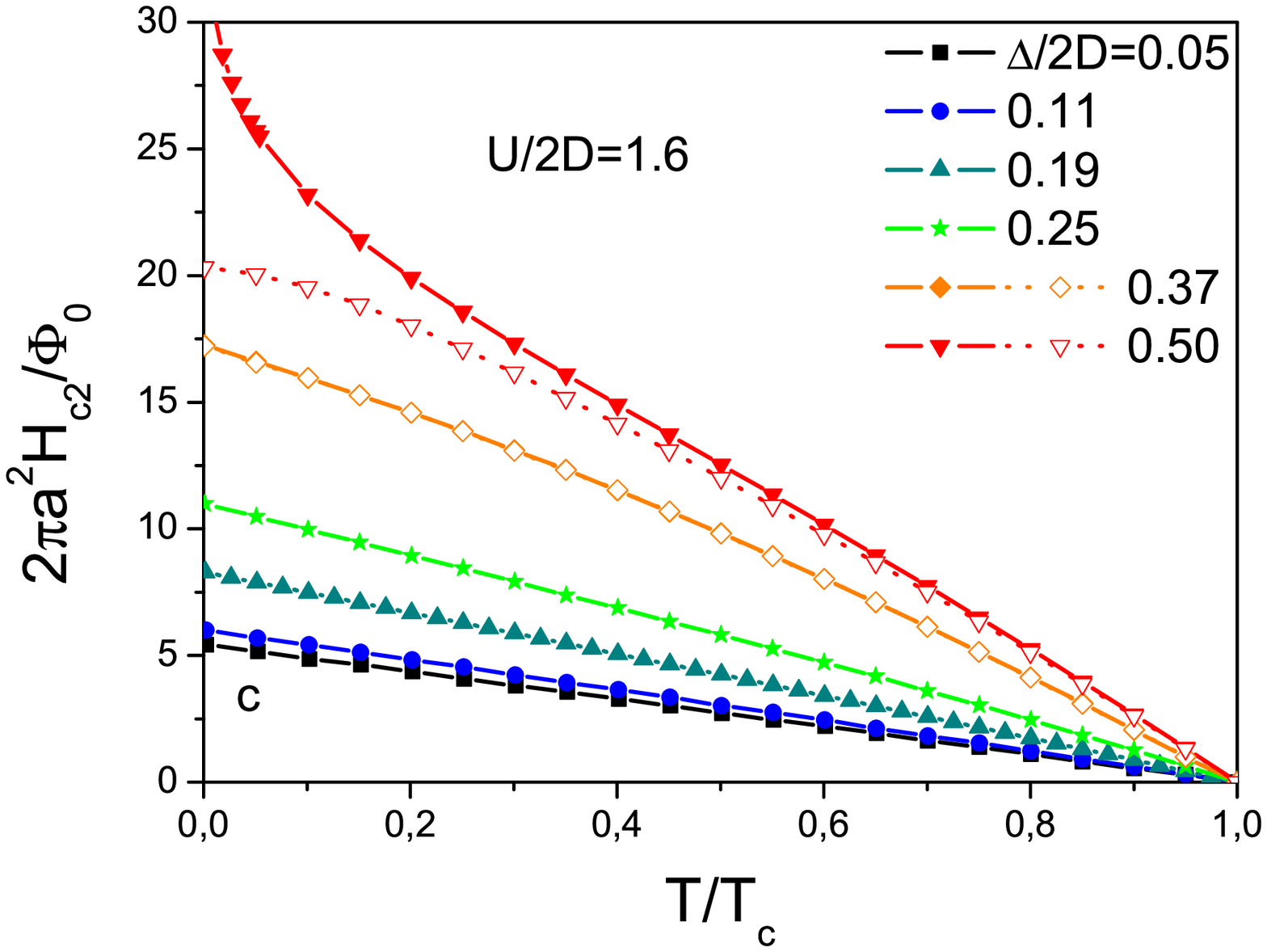}
\caption{
Temperature dependence of the upper critical field for different disorder
levels:
(a) ---  BCS weak coupling limit ($U/2D=0.2$);
(b) --- BCS -- BEC crossover region at intermediate coupling ($U/2D=1.0$);
(c) --- BEC limit of strong coupling ($U/2D=1.6$).
Filled symbols and continuous curves correspond to calculations accounting
for localization corrections. Unfilled symbols and dashed lines correspond to
the ``ladder'' approximation for impurity scattering.}
\label{fig5}
\end{figure}

In Fig. \ref{fig5} we show temperature dependencies of the upper critical field 
for different levels of disorder in three characteristic regions of coupling 
strength: in BCS weak coupling limit ($U/2D=0.2$), in BCS -- BEC crossover region 
(intermediate coupling $U/2D=1.0$) and in BEC limit of strong coupling ($U/2D=1.6$). 
In weak coupling limit (Fig.\ref{fig5}(a)) the growth of disorder leads to the
increase of the upper critical field in the whole temperature region in the
limit of weak disorder ($\Delta/2D<0.19$), while the temperature dependence has
the negative curvature and is close to the standard WHH dependence \cite{WHH}. 
With further increase of disorder with no account for localization corrections
the upper critical field decreases for all temperatures. However, taking into
account localization corrections in the weak coupling limit for the case of
strong disorder ( $\Delta/2D\geq 0.37$) significantly increases the upper
critical field and qualitatively changes its temperature dependence, so that
the curves of $H_{c2}(T)$ acquire positive curvature.
The upper critical field rapidly grows with disorder at all temperatures.
For intermediate coupling (Fig.\ref{fig5}(b)) in the limit of weak disorder 
the temperature dependence of the upper critical field becomes practically linear.
The upper critical field grows with disorder at all temperatures.
In the limit of strong disorder  ($\Delta/2D\geq 0.37$) localization corrections,
as in the weak coupling limit, increase the upper critical field  at all
temperatures and the curves of $H_{c2}(T)$ acquire positive curvature.
However, in the intermediate coupling region the influence of localization 
corrections is much weaker, than in the weak coupling limit and is relevant only
at low temperatures. In BEC limit of the strong coupling (Fig.\ref{fig5}(c)) and
in the limit of weak disorder the curves of $H_{c2}(T)$ are in fact linear.
The upper critical field grows with disorder at all temperatures. In the limit
of strong disorder at the point of Anderson transition ( $\Delta/2D=0.37$) the 
$H_{c2}(T)$ dependence remains linear and the account of localization corrections
in fact does not change the temperature dependence of the upper critical field.
Further increase of disorder leads to the increase of $H_{c2}(T)$. 
Deeply in the Anderson insulator phase ($\Delta/2D=0.5$) the  $H_{c2}(T)$ dependence
acquires the positive curvature and the account of localization effects enhances
$H_{c2}(T)$ in the low temperature region, while close to $T_c$ localization
corrections are irrelevant even at such a strong disorder.
Thus, the strong coupling significantly decreases the influence of localization
effects of the temperature dependence of the upper critical field.

\section{Conclusion}

In this paper, within the combined Nozieres -- Schmitt-Rink and $\Sigma$
generalization of the dynamical mean field theory we have investigated
the influence of disordering, in particular the strong one (including the
region of Anderson localization), and the growth of the strength of pairing  
interaction upon the temperature dependence of the upper critical field.
Calculations were performed for the wide range of attractive potentials
$U$, from the weak coupling limit of $U/2D\ll 1$, where instability of the
normal phase and superconductivity is well described by BCS model, up to
the strong coupling limit of $U/2D\gg 1$, where superconducting transition
is due to Bose -- Einstein condensation of compact Cooper pairs, which are
formed at temperatures much higher than the temperature of superconducting
transition.

The growth of the coupling strength $U$ leads to the fast increase of $H_{c2}(T)$, 
especially at low temperatures. In BEC limit and in the region of BCS -- BEC
crossover $H_{c2}(T)$ dependence becomes practically linear. Disordering also
leads to the increase of $H_{c2}(T)$ at any coupling. In the weak coupling
BCS limit the growth of disorder increases both the slope of the upper critical
field close to $T=T_{c}$ and $H_{c2}(T)$ in low temperature region.
In the limit of strong disorder in the vicinity of Anderson transition
localization corrections lead to additional sharp increase of the upper
critical field at low temperatures and $H_{c2}(T)$ dependence becomes concave, 
i.e. acquires the positive curvature. In BCS -- BEC crossover region and in
BEC limit weak disorder is insignificant for the slope of the upper critical
field at $T_{c}$, though the strong disorder in the vicinity of Anderson
transition leads to noticeable increase of the slope of the upper critical
field with the growth of disorder.
In low temperature region  $H_{c2}(T)$ significantly grows with the growth of
disorder, especially in the vicinity of Anderson transition, where localization
corrections noticeably increase  $H_{c2}(T=0)$ and  $H_{c2}(T)$ curve 
instead of linear temperature dependence, typical in the strong coupling limit
at weak disorder, becomes concave.

In our model the upper critical field at low temperatures may reach extremely 
large value significantly exceeding $\frac{\Phi_{0}}{2\pi a^2}$. This makes
important the further analysis of the model, taking into account paramagnetic
effect and inevitable role of electron spectrum quantization in magnetic field.
Actually, we can hope that effects of quantization of the spectrum are 
irrelevant in the limit of the strong disorder, while paramagnetic effect is
much weakened in the region of strong and very strong coupling. These questions
will be the task of further studies.

This work was performed within the State Contract (FASO) No. 0389-2014-0001
with partial support by RFBR grant No. 17-02-00015 and the Program of Fundamental
Research of the RAS Presidium  ``Fundamental problems of high -- temperature
superconductivity''.

\newpage


\newpage 

\begin{thebibliography}{99}
\bibitem{AG_impr}A.A. Abrikosov, L.P. Gor'kov. Zh. Eksp. Teor. Fiz. {\bf 36},
319 (1958) [Sov. Phys. JETP {\bf 9}, 220 (1959)]
\bibitem{AG_imp}A.A. Abrikosov, L.P. Gor'kov. Zh. Eksp. Teor. Fiz. {\bf 35},
1158 (1958) [Sov. Phys. JETP {\bf 9}, 1090 (1959)]
\bibitem{Gor_GL}L.P. Gor'kov. Zh. Eksp. Teor. Fiz. {\bf 36}, 1918 (1959) [Sov. Phys. JETP
{\bf 36}, 1364 (1959)]
\bibitem{AG_mimp}A.A. Abrikosov, L.P. Gor'kov. Zh. Eksp. Teor. Fiz. {\bf 39},
1781 (1960) [Sov. Phys. JETP {\bf 12}, 1243 (1961)]
\bibitem{And_th}P.W. Anderson. J. Phys. Chem. Solids {\bf 11}, 26 (1959)
\bibitem{Genn}P.G. De Gennes. Superconductivity of Metals and Alloys.
W.A. Benjamin, NY 1966
\bibitem{SCLoc_1} L.N. Bulaevskii, M.V. Sadovskii. Pis'ma Zh. Eksp. Teor. Fiz. {\bf 39}, 524
(1984) [JETP Letters {\bf 39}, 640 (1984)]
\bibitem{SCLoc_2}L.N. Bulaevskii, M.V. Sadovskii. J.Low.Temp.Phys. {\bf 59}, 89
(1985);
\bibitem{SCLoc_3}M.V. Sadovskii. Physics Reports {\bf 282}, 226 (1997);
ArXiv:cond-mat/9308018;
\bibitem{SCLoc_4}M.V. Sadovskii. Superconductivity and Localization.
World Scientific, Singapore 2000
\bibitem{WHH}N.R. Werthamer, E. Helfand. Phys.Rev. {\bf 147},  288 (1966); \\
N.R Werthamer, E. Helfand, P.C. Hohenberg. Phys.Rev.{\bf 147},  295 (1966)
\bibitem{NS} P. Nozieres and S. Schmitt-Rink, J. Low Temp. Phys. {\bf 59}, 195 (1985).
\bibitem{pruschke}  Th. Pruschke, M. Jarrell, J. K. Freericks. Adv. Phys.
{\bf 44}, 187 (1995).
\bibitem{georges96}  A. Georges, G. Kotliar, W. Krauth, M. J. Rozenberg.
Rev. Mod. Phys. {\bf 68}, 13 (1996).
\bibitem{Vollh10}D. Vollhardt in ``Lectures on the Physics of Strongly
Correlated Systems XIV'', eds. A. Avella and F. Mancini, AIP Conference
Proceedings vol. 1297 (AIP, Melville, New York, 2010), p. 339; ArXiV: 1004.5069.
\bibitem{JTL05}E.Z.Kuchinskii, I.A.Nekrasov, M.V.Sadovskii.
Pis'ma Zh. Eksp. Teor. Fiz. {\bf 82}, 217 (2005) [JETP Letters {\bf 82}, 198 (2005)];
ArXiv: cond-mat/0506215.
\bibitem{PRB05}M.V. Sadovskii, I.A. Nekrasov, E.Z. Kuchinskii, Th. Prushke,
V.I. Anisimov. Phys. Rev. B {\bf 72}, 155105 (2005);
ArXiV: cond-mat/0508585.
\bibitem{FNT06}E.Z. Kuchinskii, I.A. Nekrasov, M.V. Sadovskii. 
Fizika Nizkikh Temperatur {\bf 32}, 528 (2006) [Low Temp. Phys. {\bf 32}, 398 (2006)];
ArXiv: cond-mat/0510376.
\bibitem{UFN12}E.Z. Kuchinskii, I.A. Nekrasov, M.V. Sadovskii. Usp. Fiz. Nauk {\bf 182}, 345
(2012) [Physics Uspekhi {\bf 53}, 325 (2012)]; ArXiv:1109.2305.
\bibitem{HubDis} E.Z. Kuchinskii, I.A. Nekrasov, M.V. Sadovskii,
Zh. Eksp. Teor. Fiz. {\bf 133}, 670 (2008) [JETP {\bf 106}, 581 (2008)]; ArXiv: 0706.2618.
\bibitem{LVK16}E.Z. Kuchinskii, M.V. Sadovskii. Zh. Eksp. Teor. Fiz. {\bf 149}, 589 (2016)
[JETP {\bf 122}, 509 (2016)]; ArXiv:1507.07654
\bibitem{PRB07}E.Z. Kuchinskii, I.A. Nekrasov, M.V. Sadovskii. 
Phys. Rev. B {\bf 75}, 115102-115112 (2007); ArXiv: cond-mat/0609404.
\bibitem{JETP14}N.A. Kuleeva, E.Z. Kuchinskii, M.V. Sadovskii. Zh. Eksp. Teor. Fiz. {\bf 146},
304  (2014) [JETP {\bf 119}, 264 (2014)]; ArXiv: 1401.2295.
\bibitem{JTL14} E.Z. Kuchinskii, N.A. Kuleeva, M.V. Sadovskii. Письма ЖЭТФ
{\bf 100}, 213  (2014) [JETP Letters {\bf 100}, 192  (2014)]; ArXiv: 1406.5603.
\bibitem {JETP15} E.Z. Kuchinskii, N.A. Kuleeva, M.V. Sadovskii. Zh. Eksp. Teor. Fiz. {\bf 147},
1220  (2015) [JETP {\bf 120}, 1055 (2015)]; ArXiv:1411.1547.
\bibitem {JETP16} E.Z. Kuchinskii, N.A. Kuleeva, M.V. Sadovskii. Zh. Eksp. Teor. Fiz. {\bf 149},
430 (2016) [JETP {\bf 122} 375 (2016)]; ArXiv:1507.07649.
\bibitem {FNT16} E.Z. Kuchinskii, N.A. Kuleeva, M.V. Sadovskii. Fizika Nizkikh Temperatur {\bf 43},
22 (2017) [Low Temp. Phys. {\bf 42}, 17  (2017)];
ArXiv: 1606.05125.
\bibitem{JETP17} E.Z. Kuchinskii, N.A. Kuleeva, M.V. Sadovskii.  
Zh. Eksp. Teor. Fiz. {\bf 152}, 133 (2017)[JETP {\bf 125}, 111 (2017)]; ArXiv:1702.05247
\bibitem{NRGrev} R. Bulla, T.A. Costi, T. Pruschke, Rev. Mod. Phys. {\bf 60}, 395 (2008).
\bibitem{LP}E.M. Lifshits, L.P. Pitaevskii. Statisticheskaya Fizika. Part 2.
Ch. 5, ``Nauka'', Moscow 1978 [E.M. Lifshits, L.P. Pitaevskii. Statistical
Physics. Part 2. Ch. 5, Pergamon Press, Oxford, 1980]
\bibitem{Hc2loc_d3}E.Z. Kuchinskii, M.V. Sadovskii. Sverkhprovodimast': Fizika, Chimija, Tekhnika 
{\bf 4}, 2278 (1991)
[Superconductivity: Physics, Chemistry, Technology {\bf 4}, 2278 (1991)]
\bibitem{Hc2loc_d3a}E.Z. Kuchinskii, M.V. Sadovskii. Physica C{\bf 185-189},
1477 (1991)
\bibitem{AGD}A.A. Abrikosov, L.P. Gor'kov, I.E. Dzyaloshinskii. Metodi kvantovoi
teorii polja v statisticheskoi fizike. Fizmatgiz, Moscow, 1963. [A.A. Abrikosov,
L.P. Gor'kov, I.E. Dzyaloshinskii. Quantum Field Theoretical Methods in
Statistical Physics. Pergamon Press, Oxford, 1965]
\bibitem{Diagr}M.V. Sadovskii. Diagrammatika. Moscow -- Izhevsk, 2010;
M.V. Sadovskii. {\em Diagrammatics}. World Scientific, Singapore 2006; .
\bibitem{VW}D.~Vollhardt and P.~W\"olfle, Phys. Rev. B {\bf 22}, 4666 (1980);\
Phys. Rev. Lett. {\bf 48}, 699 (1982).
\bibitem{WV}P.~W\"olfle and D.~Vollhardt, in {\em Anderson Localization},
eds. Y.~Nagaoka and H.~Fukuyama, Springer Series in Solid State Sciences,
vol. 39, p.26. Springer Verlag, Berlin 1982.
\bibitem{MS}A.V. Myasnikov, M.V. Sadovskii, Fizika Tverdogo Tela {\bf 24}, 3569 (1982)
[Sov. Phys.-Solid State {\bf 24}, 2033 (1982) ].
\bibitem{KS} E.A.~Kotov, M.V.~Sadovskii. Zs. Phys. B {\bf 51}, 17 (1983).
\bibitem{MS86}M.V.~Sadovskii, in {\em Soviet Scientific Reviews -- Physics
Reviews}, ed. I.M.~Khalatnikov, vol. 7, p.1. Harwood Academic Publ., NY 1986.
\bibitem{VW92}D.~Vollhardt, P.~W\"olfle, in {\em Electronic Phase Transitions},
eds. W.~Hanke and Yu.V.~Kopaev, vol. 32, p. 1. North--Holland, Amsterdam 1992.

\end{thebibliography}
\end{document}